\documentstyle[12pt,psfig,oldlfont]{article}

\font\tenrm=cmr10

\begin{document}

\renewenvironment{thebibliography}[1]
  { \begin{list}{\arabic{enumi}.}
    {\usecounter{enumi} \setlength{\parsep}{0pt}
     \setlength{\itemsep}{3pt} \settowidth{\labelwidth}{#1.}
     \sloppy
    }}{\end{list}}

\parindent=1.5pc

\renewcommand{\thefootnote}{\fnsymbol{footnote} }

\newcommand{\s}{\\ \vspace*{-2mm} }
\newcommand{\nn}{\noindent}
\newcommand{\non}{\nonumber}
\newcommand{\ee}{e^+ e^-}
\newcommand{\ra}{\rightarrow}
\newcommand{\lra}{\longrightarrow}
\newcommand{\beq}{\begin{eqnarray}}
\newcommand{\eeq}{\end{eqnarray}}
\newcommand{\tb}{{\rm tg} \beta}
\newcommand{\SMG}{{\rm SU(3)_C \times SU(2)_L \times U(1)_Y}}
\newcommand{\lsim}{\raisebox{-0.13cm}{~\shortstack{$<$ \\[-0.07cm] $\sim$}}~}
\newcommand{\gsim}{\raisebox{-0.13cm}{~\shortstack{$>$ \\[-0.07cm] $\sim$}}~}

\begin{flushright}
KA--TP--13--95\\
December 1995 \\
\end{flushright}

\begin{center}{{\bf EXTENDED GAUGE MODELS AT e$^+$e$^-$ COLLIDERS\footnote{
Invited talk given at the Workshop on Physics and Experiments with
Linear Colliders, Morioka--Appi, Japan, September 8--12 1995.}}
\vglue .4cm
{A. DJOUADI}\\
\baselineskip=14pt
{\it Institute f\"ur Theoretische Physik, Universtit\"at Karlsruhe,}\\
\baselineskip=14pt
{\it D--76128 Karlsruhe, FRG}.\footnote{Present address. \\ }\\
\vglue 0.1cm
and
\vglue 0.1cm
{\it Groupe de Physique des Particules, Universit\'e de Montr\'eal,}\\
\baselineskip=14pt
{\it Case 6128A, H3C 3J7 Montr\'eal, Canada.}\\
\vglue 0.5cm
{\tenrm ABSTRACT}}
\end{center}
{\rightskip=3pc
 \leftskip=3pc
\tenrm\baselineskip=12pt
\noindent
We summarize the potential of high--energy $\ee$ linear colliders for
discovering, and in case of discovery, for studying the signals of
extended gauge models. We will mainly focus on the virtual signals of
new neutral gauge bosons and on the production of new heavy leptons.
\vglue 0.6cm}

%

{\bf\noindent 1. Introduction}
\vglue 0.2cm
\baselineskip=14pt

Despite of its tremendous success in describing the experimental data
within the range of energies available today, the Standard Model (SM)
based on the gauge symmetry $\SMG$ is widely believed not to be the
ultimate truth. Besides the fact that it has too many parameters which
are incorporated by hand, the SM does not unify the electroweak and
strong forces in a satisfactory way since the coupling constants of
these interactions are different and appear to be independent.
Therefore one would expect that a more fundamental theory exists which
describes the three forces within the context of a single gauge group
[which will contain ${\rm SU(3) \times SU(2)}$ $\times$ U(1) as a
subgroup and will reduce to this symmetry at low energies] and hence,
with only one coupling constant \cite{S1,S2}. Recent LEP data show that
this can be indeed achieved in Supersymmetric Grand Unified Theories
\cite{S3}.

\smallskip

Two predictions of Grand Unified Theories can have dramatic
phenomenological consequences in the ${\cal O}$(TeV) energy range:

\smallskip

{\it (i)} The unifying group must be spontaneously broken at the
unification scale, $\Lambda_{\rm GUT} \sim 10^{16}$~GeV in order to be
compatible with the experimental bounds on the proton lifetime. However,
it is possible that the breaking to the SM group occurs in several steps
and that some subgroups remain unbroken down to a scale of order 1 TeV.
In this case the surviving group factors allow for new gauge bosons with
masses not far from the scale of electroweak symmetry breaking.

{\it ii)} The grand unified groups incorporate fermion
representations	in which a complete generation of SM quarks and
leptons can	be naturally embedded.  In most of the cases these
representations	are large enough to accomodate additional new
fermions which are needed to have anomaly--free theories.  It is
conceivable	that these new fermions [for instance, if they are
protected by symmetries] acquire masses not much larger than the
Fermi	scale. This is	very likely, and even necessary if the predicted new
gauge bosons are relatively light \cite{S4}.

\smallskip

Besides the SU(5) group [the simplest Lie group containing
SU(3)$\times$SU(2) $ \times $U(1) as a subgroup and two representations
to accomodate the 15 SM fermions], which has no room for relatively
light new gauge bosons or new fermions, two other unifying groups have
received much attention in recent years, SO(10)	\cite{S5} and the
exceptional group E$_6$ \cite{S6}.

SO(10) is the simplest group in which the 15 Weyl spinors of
each SM generation of fermions can be embedded into a single
multiplet.  This representation	has dimension {\bf 16} and, in order to be
anomaly--free, contains a right--handed neutrino
\begin{eqnarray}
{\rm \left[ \begin{array}{c}	\nu_e \\ e \end{array}
\right]_L
\hspace{0.3cm} \begin{array}{c}	\\ e_R  \end{array} \hspace{0.3cm}
\left[ \begin{array}{c}	u	\\ d \end{array} \right]_L
\hspace{0.3cm}
\begin{array}{c} u_R \\ d_R	\end{array} \ \	;
\hspace{0.5cm} \nu_{eR} }
\end{eqnarray}
\nn The	gauge group can	be spontaneously broken to the SM group at an
intermediate scale,  two interesting chains of breaking	patterns being via
SU(5)$ \times$U(1) or SU(4)$ \times$ SU(2)$\times$SU(2), leading to the
intermediate symmetries \cite{S7}
\begin{small}
\begin{eqnarray}
{\rm SO(10)} \rightarrow & \SMG	\times {\rm U(1)_\chi} & \rightarrow \SMG
\nonumber \\
\rightarrow & {\rm SU(3)_C \times SU(2)_L \times SU(2)_R \times
U(1)_{B-L}} & \rightarrow \SMG \hspace*{1cm}
\end{eqnarray}
\end{small}
These chains would induce new right--handed charged currents and/or
neutral currents which could eventually be studied at TeV energies.  The
most general $Z'_{LR}$ would couple to the current
$J^\mu_{LR}=\alpha_{LR} J_{3R}^\mu - J^\mu_{B-L}/(2\alpha_{LR})$ where the
parameter $\alpha_{LR}$ is defined in terms of the SU(2)$_L$ and
SU(2)$_R$ couplings as $ \alpha_{LR} \equiv [ (g_R
\cos\theta_W)^2/(g_L \sin \theta_W)^2$ $-1 ]^{1/2}$ and lies in the
range $\sqrt{2/3} \leq \alpha_{LR} \leq \sqrt{2}$.

Another popular unifying group is E$_6$.  It contains SU(5) and SO(10) as
subgroups and is the next anomaly--free choice after SO(10).  The
interest in this group is mainly due the fact that superstring theories,
which attempt to unify all fundamental forces including gravity, suggest this
theory as a possible four dimensional field theoretic limit \cite{S8}.
[More recently, some alternative scenarios appeared, though.] In
E$_6$, each quark--lepton generation lies in a representation of dimension
{\bf 27}. To complete this representation, twelve new fields are needed
in addition to the SM fermion fields. For each family one has two additional
isodoublets of leptons, two isosinglet	neutrinos and an isosinglet quark
with charge $-1/3$,
\begin{eqnarray}
{\rm \left[ \begin{array}{c}	\nu_E \\ E \end{array}
\right]_L
\hspace{0.3cm}
\left[ \begin{array}{c}	\nu_E \\ E \end{array}
\right]_R
\hspace{0.3cm}
\nu_{eR} 	\hspace{0.3cm} n
\hspace{0.6cm} D_L \hspace{0.3cm}	D_R }
\end{eqnarray}

Moreover, the supersymmetric partners of these fermions will lie in the
representation of dimension $\overline{\bf 27}$; the third generation
partners of doublets and singlet lepton fields will acquire vacuum
expectation values, providing the Higgs sector of the theory.

In the breaking of E$_6$ down to the SM gauge group, two additional
U(1)	symmetry factors may survive at low energies \cite{S6}
\begin{small}
\begin{eqnarray}
{\rm E_6} \rightarrow {\rm SO(10) \times U(1)_\psi} & \rightarrow
& {\rm SU(5)
\times U(1)_\chi \times	U(1)_\psi} \\ \nonumber	& \rightarrow &
\SMG \times {\rm U(1)_\chi \times U(1)_\psi}
\end{eqnarray}
\end{small}

\hspace*{-0.52cm}leading to two new neutral gauge bosons. Assuming that only
one of them is relatively
light, the relevant neutral gauge boson would be  $Z' = Z_\chi
\cos \beta + Z_\psi	\sin \beta$, where $\beta$=0	and $\pi/2$ correspond
to pure Z$_\chi$ and	Z$_\psi$ while $\beta={\rm arctg} (-\sqrt{5/3})$
corresponds to the model $\eta$	in which E$_6$ is directly broken to a	rank--5
group at the unification scale in superstrings models \cite{S8}.

Several other gauge groups have been considered, based on various
theoretical motivations.  For instance,	schemes	of grand unification built--up
on large orthogonal groups have been proposed to explain the origin of parity
violation in weak interactions	\cite{S9}. In	these models,	weak
interactions are $P$--invariant but fermions with left--handed and
right--handed couplings acquire	different masses.  They predict	a new
spectrum	of fermions, mirror fermions \cite{S10}, which have
chiral properties opposite to the ordinary particles. In the simplest version
of these models, the gauge symmetry and the symmetry breaking pattern are
the same as in the SM; three families of heavy fermions with
opposite chiralities are simply added to the SM spectrum
\begin{eqnarray}
{\rm
\left[ \begin{array}{c}	N_R  \\ E_R	\end{array} \right]
\hspace{0.3cm}
       \begin{array}{c}	 N_L  \\ E_L \end{array} \hspace{0.6cm}
\left[ \begin{array}{c}	U_R \\ D_R \end{array} \right]
\hspace{0.3cm}
       \begin{array}{c}	U_L \\ D_L \end{array} }
\end{eqnarray}
\nn Theoretical	arguments based	on [weak coupling] unitarity \cite{S11}
suggest that the masses of these mirror fermions should not exceed a
few hundred GeV.

In many extensions of the SM, particles with exotic quantum numbers are
also predicted to occur \cite{S12}. For instance Leptoquarks [which have baryon
and
lepton numbers B=$\pm$1/3 and L=$\pm$1 and couple to quark--lepton
pairs] are generic predictions of GUTs and point to the necessity of
unifying the strong and the electroweak forces. Diquarks [B=$\pm$2/3 and
L=0] and Dileptons [B=0 and L=$\pm$2] are also predicted by some gauge
extensions of the SM. Masses for Difermions [Leptoquarks, Diquarks and
Dileptons] below the TeV scale are still compatible with present
experimental bounds.

The direct search for these new matter and gauge particles and tests of
their indirect effects will be a major goal of the next generation of
accelerators. In this talk, I will summarize the potential of
high--energy $\ee$ linear colliders for these searches, focussing on the
virtual effects of new neutral gauge bosons [only the $\ee$ option will
be considered, $Z'$ effects can also be searched for in $e^- e^-$
scattering \cite{S13} but this is just equivalent to the search in
Bhabha scattering] and on the production and the study of new heavy
fermions and difermions.

\newpage

{\bf\noindent 2. New Gauge Bosons}
\vglue 0.2cm
\baselineskip=14pt

{\it\noindent 2.1 Physical Set--Up}
\vglue 0.2cm
\baselineskip=14pt

We will concentrate on the two most theoretically motivated effective
theories which lead to an additional neutral gauge boson $Z'$: SU(2)$_L
\times$U(1)$_Y \times$U(1)$_{Y'}$ originating from the breaking of E$_6$
and Left-Right (LR) models based on the symmetry SU(2)$_L \times$SU(2)$
_R \times$U(1). A discussion of alternative models can be found in Ref.
\cite{R6}.

Once the hypercharge of the additional U(1) factor in E$_6$ models [we
will assume as usual the GUT relation $g_Y=g_{Y'}$] and the coupling of
the SU(2)$_R$ group [or alternatively $\alpha_{LR}$] in LR models are
fixed, the couplings of the $Z'$ boson to fermions is uniquely
determined. These gauge couplings will be altered by $Z$--$Z'$ mixing,
though. Indeed, the physical [mass] eigenstates $Z$ and $Z'$ are
admixture of the weak eigenstates with a mixing angle $\theta_{\rm
mix}$ [for sufficiently large $Z'$ masses] given by $\theta_{\rm mix} =
P M_Z^2/M_{Z'}^2$, where the parameter $P$ depends on the symmetry
breaking pattern. However, this mixing will also alter the couplings of
the fermions to the $Z$ boson which have been very accurately measured
at LEP1 and found to be in a very good agreement with SM expectations.
Several analyses of LEP1 data show that this angle is smaller than
$\theta \lsim 0.005$ \cite{R1} and it can be safely neglected.

The bound on $\theta_{\rm mix}$ directly translates into a lower bound
on the $Z'$ masses. Since in the theoretically well motivated models, the
coefficient $P$ is of order unity, one is led to a lower bound of the
order of several hundred GeV on $M_{Z'}$ [especially in the
``constrained" case where the Higgs sector is specified]. Stringent
limits are also available from a negative [direct] search of a peak in
the invariant mass spectrum of $\ee$ pairs  at the Tevatron. Depending
on the models, masses up to 450 GeV are already excluded by both direct
and indirect searches; Tab.1 \cite{R2}.

\begin{center}
\begin{tabular}{|c||c|c|c|} \hline
\hspace*{0.2cm} Model \hspace*{0.2cm} & \hspace*{0.2cm} Direct
\hspace*{0.2cm} & \hspace*{0.2cm} Indirect  \hspace*{0.2cm} &
\hspace*{0.2cm} Indirect \hspace*{0.2cm} \\
               & Tevatron & unconstrained &constrained \\ \hline
      $\chi$   & 425      &  330          &  920       \\
      $\psi$   & 415      &  170          &  170       \\
      $\eta$   & 440      &  220          &  610       \\
      LR       & 445  &  390          &  1360       \\ \hline
\end{tabular}
\end{center}
\vspace{0.2cm}

\noindent{\small {\bf Tab.1}: Present constraints on $M_{Z'}$ [in GeV]
from direct
and indirect searches  for the $E_6$ models $\chi, \eta, \psi$ and for a
LR model with $g_L=g_R$; from Ref.\cite{R2}.}
\vspace*{.4cm}

The existence of an extra neutral gauge boson with a mass below the
maximal energy of and $\ee$ collider will provide a new	resonance
which	will increase the $\ee$ annihilation cross section by several
orders of magnitude. In this case, an $\ee$ collider operating at the
resonance peak would be a ``$Z'$ factory" allowing to measure the
couplings of the $Z'$ to other conventional and new particles with very
high precision, a situation comparable to the LEP experiments exploring
the $Z$ peak. High--energy $\ee$ colliders would then be ideal
instruments to study the properties of the new gauge bosons and to
constrain the theories predicting their origin.

Even if a new vector boson is too heavy to be produced as a resonance,
it could give rise to virtual effects which are measurable. Indeed,
besides mixing with the $Z$, the $Z'$ will participate in the
production process of ordinary fermions [15--18]
\vspace*{-2mm}
\begin{eqnarray}
\ee \longrightarrow \gamma, Z, Z' 	\longrightarrow	f \bar{f}
\end{eqnarray}
and will affect the cross sections and the various asymmetries
through	its propagator effects. This situation will be similar to PEP,
PETRA and TRISTAN seeing the $Z$ propagator effects at relatively low
energies. The clean environment of $\ee$ colliders allows to probe
these virtual effects with high precision and therefore provides a
sensitivity to $Z'$ masses considerably higher than the available c.m.
energy. In addition, because of the number of observables which can be
measured precisely, a detailed investigation of the $Z'$ properties can
be performed and its origin can be identified.

In the following we will assume that the $Z'$ is heavier than the c.m.
energy, and study its propagator effects on various observables of the
process $\ee\rightarrow f\bar{f}$. We wil not discuss $Z'$ searches in
the process $\ee \ra WW$: since the $Z'WW$ coupling is generated only
through $Z$--$Z'$ mixing in the models we are considering [see
Ref.\cite{R6} for alternative models] the effects are very small
\cite{R6a}.

\bigskip

{\it\noindent 2.2 Signals and backgrounds}
\vglue 0.2cm
\baselineskip=14pt

Since the effects of new gauge bosons with masses far beyond the
production threshold are expected to be rather small, one has to take
into account radiative corrections and especially initial state QED
radiation which gives very large contributions. Indeed, the radiative
tail of the $Z$ boson enhances the SM cross section by a factor 2 to 3
at a c.m. energy of $\sqrt{s}=500$ GeV, thus completely diluting the
expected ${\cal O}$ (few percent) $Z'$ signals \cite{R5}.

At high energies, a radiative tail arises if the cut $\Delta$ on the
photon energy is such that $\Delta = E_\gamma/E_{\rm beam}> 1-M_Z^2/s$.
In order to remove the tail, one obviously has to choose a cut $\Delta<
1-M_Z^2/s$, which at 500 GeV for instance, would be the case for
$\Delta<0.967$ i.e. $E_\gamma <242$ GeV. Then, the ratio of the $Z'$
signal to the $Z$ contribution remains of comparable size as it was at
the Born level, although it has a different numerical value. This is
summarized in Fig.~1a, where the ratio of the hadronic cross section
[with the five light quarks] to the muonic cross section is plotted at
$\sqrt{s}=500$ GeV,  assuming a $Z'$ [in the model $\chi$ which is
equivalent to a LR model with $\alpha_{LR}=\sqrt{2/3}$] with
$M_{Z'}=750$ GeV. One sees that the cut on $E_\gamma$ strongly
influences not only the cross section values, but also the difference
between observables with and without $Z'$ exchange.

Another important point is that the range of $Z'$ masses and couplings
which can be probed directly depends on the experimental errors with
which the physical observables are measured. In order to make realistic
estimates of these errors and select those observables which will be
measured with the best accuracy, the experimental situation should be
taken into account.

At high energies the hadronic cross section is several orders of
magnitude smaller than at LEP and comparable or even smaller than the
$WW$ cross section. Harder cuts should be therefore applied against the
two--photon backgrounds which rises like log$s$. Fig.1b shows the energy
spectrum [displayed as a function of $\Delta$] for the dominant
processes with hadronic final states at $\sqrt{s}=500$ GeV. To suppress
the $\gamma \gamma$ background, a conservative cut $\Delta <0.7$
[also needed to remove the $Z$ tail] for which typical event rates for a
luminosity $\int {\cal L}=20$ fb$^{-1}$ are: $1.6 \times 10^{4}$ for
leptons [with $s$--channel exchange for $\ee$ only] and $5.2 \times 10^{4}$
for hadrons [without top quarks].

\vspace*{7cm}

\noindent{\small {\bf Fig.1}: a) The ratio $\sigma^{\rm had}/\sigma^{\rm
lept}$ as a function of $s$ with/without: QED radiative corrections, a cut on
the photon energy and $Z'$ exchange. (b) Final state energy spectrum
[expressed in terms of $\Delta=1-s'/s$, with $s'$ is the ``measured"
sum] of the dominant process with hadronic final states at $\sqrt{s}=
500$ GeV; initial state radiative is included but not beamstrahlung. }
\vspace*{.2cm}

For the lepton detection, similar systematic errors as at LEP can be
achieved, but for $\ee \ra q\bar{q}$ the situation is different because
of the $WW$ background and one has to attribute an error of $\sim 1\%$
to the selection efficiency. An absolute luminosity error of $1\%$ is
feasible if beamstrahlung effects are under control. The estimates for
statistical and systematic errors suggest to base the analysis on the
variables
\begin{eqnarray}
\sigma^{\rm lept} \ , \ R= \sigma^{\rm had}/\sigma^{\rm lept} \ , \
A_{\rm FB}^{\rm lept}
\end{eqnarray}
and if longitudinal polarization is available
\begin{eqnarray}
A_{\rm LR}^{\rm lept} \ , \  A_{\rm LR}^{\rm had}
\end{eqnarray}
with the expected systematic errors summarized in Tab.2. [Note that if
efficient $b$ tagging is also available, one can use $R_b=\sigma^b/
\sigma^{\rm had}, A_{\rm FB}^b$ and $A_{\rm LR}^b$ to obtain additional
information \cite{R2,R7}; a detailed analysis of this final state is under
way \cite{R7}.]
\vspace{0.1cm}

\begin{center}
\begin{tabular}{|c||c|c|c|c|c|} \hline
   & $ \ \Delta \epsilon_\mu / \epsilon_\mu \  $
   & $ \ \Delta \epsilon_{\rm had} / \epsilon_{\rm had} \ $
   & $ \ \Delta A_{\rm FB}^{\rm lept} \ $
   & $ \ \Delta A_{\rm LR}  \ $
   & $ \ \Delta {\cal L} / {\cal L} \ $ \\ \hline
\ sys. err. \ & 0.5\% & 1\% & negl. & 0.003 & 1\% \\ \hline
\end{tabular}
\vspace{0.3cm}

\noindent{\small {\bf Tab.2}: Systematic errors on the observables
eqs.(7--8) at $\sqrt{s}=500$ GeV.}
\end{center}

\bigskip

{\it\noindent 2.3 $Z'$ mass reach and identification}
\vglue 0.2cm
\baselineskip=14pt

By comparing the measurement of the previous observables with the
predictions of various models one can derive the mass of the $Z'$
which can be probed. If all observations happen to be consistent
with SM predictions, the lower bounds on $M_{Z'}$ for $E_6$ and LR
models are shown at the 95\% confidence level as functions of
$\cos \beta$ and $\alpha_{LR}$ in Figs.2--3.

\vspace*{6cm}

\noindent{\small {\bf Fig.2}: 95\% CL $Z'$ mass reach in $E_6$ and LR
models as functions of $\cos\beta$ and $\alpha_{LR}$ respectively, when
combining the measurements of $\sigma^{\rm lept}$, $R$ and $A_{\rm
FB}^{\rm lept}$ with [thick solid curve] and without [thick dotted
curve] systematic errors. The thin lines are the limits when longitudinal
polarization is included; from \cite{R5}.}

\vspace*{6cm}

\noindent{\small {\bf Fig.3}: 95\% CL $Z'$ mass reach in $E_6$ and LR
models for a c.m. energy of 0.5 TeV [solid lines], 1 TeV [dashed lines]
and 2 TeV [dotted lines], with [thin lines] and without [thick lines]
polarization. A fixed luminosity of 20 fb$^{-1}$ and the errors in
Tab.2 are assumed.}
\vspace*{.2cm}

In Fig.2, the $Z'$ mass limits are shown for a c.m. energy of 500 GeV
and a luminosity $\int {\cal L}=20$ fb$^{-1}$ when combining all the
measurements eq.(7--8). In these figures, the leading radiative
corrections [see \cite{R7} for the full electroweak corrections] a cut
$\Delta <0.7$ and the photon energy and the systematics of Tab.2 are
taken into account.
To demonstrate the effects of systematic errors, the limits calculated
from statistical errors only are shown. Also are shown the limits
obtained by using longitudinal polarization. Without polarization, $Z'$
masses up to 3 TeV can be probed for some parameter values; for LR
models longitudinal polarization would allow to extend the mass reach by
up to 1 TeV. In the limit of vanishing statistics, the $Z'$ mass reach
can be pushed up to 4 TeV for certain parameters.

Fig.3 shows how these limits scale with the c.m. energy, keeping
systematic errors as in Tab.2 and a \underline{constant} luminosity of
20 fb$^{-1}$. Raising the c.m. energy to 1 TeV or 2 TeV allows to extend
the $Z'$ mass reach to 5 TeV and 8 TeV respectively. At these energies
most of the uncertainties are due to the statistics [the rates scale
like $1/s$]; at 2 TeV collider, collecting an integrated luminosity of
300 fb$^{-1}$ will raise these limits by approximately a factor 2. This
means that $Z'$ masses up to 16 TeV could be reached at a 2 TeV $\ee$
collider for certain model parameters.

\vspace*{7cm}

\noindent{\small {\bf Fig.4}: Left: distinction between $E_6$ and LR
models for $M_{Z'}=1.5$ (a) and 2 TeV. Right: determination
of the $E_6$ parameter $\cos\beta$ for $M_{Z'}=1$ (a) and 2 TeV (b).
The c.m. energy is $\sqrt{s}=500$ GeV and the luminosity is 20
fb$^{-1}$. The significance of the measurement of the observables
in eq.(7) [hatched area] and eq.(8) [cross--hatched area] have been
combined. The results are at the 95\%CL; from \cite{R5}.
}
\vspace*{0.3cm}

If a $Z'$ signal has been observed, the next step would be to
try to elucidate its model origin. Assuming that the observations are
consistent with the $E_6$ prediction and that $M_{Z'}$ is
determined [a model independent approach will be discussed in the next
section], Fig.4a shows which values of the LR model parameter
$\alpha_{LR}$ can be excluded at the 95\% CL at $\sqrt{s}=500$ GeV
for $M_{Z'}=1.5$ and 2 TeV. The regions of
confusion are the hatched [without polarization] and cross--hatched
[with polarization] areas. As can be seen, the role of the longitudinal
polarization is crucial: for $M_{Z'}=1.5$ TeV [for 1 TeV, there is
almost no confusion except at the point $\alpha_{LR}=\sqrt{2/3}$
corresponding to model $\chi$], only the small cross-hatched area does
not allow the distinction between $E_6$ and LR models. This
confusion becomes larger with increasing $Z'$ mass, but even for
$M_{Z'}=2.5$ TeV the distinction is still possible.

In a given class of models, the determination of the parameter
itself is also possible. This is shown in Fig.~4b for $E_6$
and two $Z'$ mass values 1 and 2 TeV: given that the data are
consistent with a certain value of $\cos \beta$ indicated on the
abscissa, the 1$\sigma$ range for $\cos \beta$ is shown in ordinate.
Here again, the gain in sensitivity if longitudinal polarization is
available is very important, and even for a $Z'$ mass of 2.5 TeV,
$\cos \beta$ can be constrained. Similar results can be obtained for
LR models.

\bigskip

{\it\noindent 2.4 Model independent approach}
\vglue 0.2cm
\baselineskip=14pt

In order not to commit oneself to a particular model, the search for $Z'$
effects in a model independent way is more adequate; in this case, the
search would also include new gauge bosons originating from alternative
models than $E_6$ and LR. A model independent analysis of $Z'$ signals
is being performed by Leike and Riemann \cite{R7}; I will summarize the main
points below.

Assuming that a new $Z'$ is present, its effect in the $\ee \ra
f\bar{f}$ process will be to add an extra contribution to the amplitude
that one can write in the general case
\begin{eqnarray}
{\cal M} (Z')= g_{Z'}^2/(s-M_{Z'}^2) \
\bar{v}_e \gamma_\alpha (v_e'+\gamma_5 a_e') u_e \
\bar{u}_f \gamma^\alpha (v_f'+\gamma_5 a_f') v_f
\end{eqnarray}
where $v'_f$ and $a'_f$ [and $g_{Z'}$] are kept free; this contribution
can be rewritten as
\begin{eqnarray}
{\cal M} (Z')= -(4\pi/s) \ \bar{v}_e \gamma_\alpha (V_e^N+\gamma_5
A_e^N) u_e \ \bar{u}_f \gamma^\alpha (V_f^N+\gamma_5 A_f^N) v_f
\end{eqnarray}
Far below the $Z'$ resonance, only $V_f^N$ and $A_f^N$ given by
\begin{eqnarray}
V_f^N= v_f' \sqrt{g_{Z'}^2/4\pi \times s/(M_{Z'}^2-s)} \ \ , \ \
A_f^N= a_f' \sqrt{g_{Z'}^2/4\pi \times s/(M_{Z'}^2-s)}
\end{eqnarray}
can be constrained and not $a_f', v_f', g_{Z'}$ and $M_{Z'}$ separately.
In the case of the leptonic observables of eqs.(7--8) [assuming
universality]
the signal of a $Z'$ is seen if \cite{R7a}
\begin{eqnarray}
\sigma^{\rm lept} \  & {\rm if} & \ (V_e^N/H^V_1)^2+ (A_e^N/H^A_1)^2
\geq 1  \non \\
A_{\rm FB}^{\rm lept} \  & {\rm if} & \ (V_e^N/H^V_2)^2-
(A_e^N/H^A_2)^2 \geq 1 \non \\
A_{\rm LR}^{\rm lept} \  & {\rm if} & \ (V_e^N/H^V_3) \times
(A_e^N/H^A_3) \geq 1
\end{eqnarray}
with the factors $H_{1,2,3}^{V,A}$ depending on the errors on the
observables; if only statistical errors are taken into account, one
would have
\begin{eqnarray}
H^{V,A}_{1,2,3} \sim \sqrt{\Delta \sigma/\sigma} \ , \ \Delta A_{\rm FB}
\ , \Delta A_{\rm LR} \ \sim [ \int {\cal L}/s ]^{-1/4}
\end{eqnarray}
The limits on the $Z'$ mass that can be obtained in this case will scale
as
\begin{eqnarray}
M_{Z'}^{\rm max} \sim (\sqrt{s}/ V_e^N) \ , \ (\sqrt{s}/ A_e^N) \
\sim [ \int {\cal L}/s ]^{1/4}
\end{eqnarray}
The analysis of Ref.\cite{R5} has been repeated within this approach
[with, in addition, the inclusion of the full electroweak corrections].
Assuming an energy of 500 GeV and a luminosity of 20 fb$^{-1}$, and
allowing from slightly worse systematical errors than in Ref.\cite{R5}
as well as a polarization degree of 60\%, a constraint $(V_e^N)^2 +
(A_e^N)^2 \lsim 0.01$ has been obtained in the case where the combined
measurement of the three lepton observables is in accord with the SM
prediction. Once this constraint is available, one just need to specify
the couplings $v_e',a_e'$ and $g_{Z'}$ to derive the limit on the $Z'$
mass in a given model.

The analysis has also been repeated in the case of $b$--quark final
states but the limits which have been obtained are worse a consequence
of the larger systematic errors for $b$ final states and the presence of
the additional $Z'bb$--couplings in the fits. However, this particular
final state gives new information in special cases.

\smallskip

The unambiguous determination of the $Z'$ parameters [once its signals
have been discovered] in a model independent way is a much
harder task. For models of extended gauge origin, a method for the
determination of the gauge couplings and the reconstruction of the gauge
structure of the $Z'$ in a model independent way [the reconstruction
of the symmetry breaking pattern is much more complicated since
$Z$-$Z'$ mixing is small] has been recently proposed by del Aguila,
Cveti\~c and Langacker \cite{R8}:

In the general case, the $Z'$ couplings are specified by five charges
[for left-- and right--handed quarks and leptons], the overall
strength $g_2$ and the coupling $g_{12}$ to the weak hypercharge [these
couplings are fixed at the GUT scale, but are renormalized at low
energies]; in addition, there is an eighth parameter which is $M_{Z'}$.
In $\ee$ collisions, one can probe four independent normalized
charges which are defined as
\begin{eqnarray}
P_V^e= \frac{g_{L2}^e+g_{R2}^e}{g_{L2}^e-g_{R2}^e} \ , \
P_L^q= \frac{g_{L2}^q}{g_{L2}^e-g_{R2}^e} \ , \
P_R^u= \frac{g_{R2}^u}{g_{L2}^q} \ , \
P_R^d= \frac{g_{R2}^d}{g_{L2}^q}
\end{eqnarray}
as well as the ratio involving $g_2$ and the $Z'$ propagator
\begin{eqnarray}
\epsilon_A = (g_{L2}^e-g_{R2}^e)^2 (s g_2^2)/(4\pi \alpha)(M_{Z'}^2-s)
\end{eqnarray}
A 500 GeV collider with $\int {\cal L}=20$ fb$^{-1}$, when including only
statistical errors and assuming 100\% longitudinal polarization and
100\% $b$ and $t$--tag efficiencies, allows the determination of the five
parameters eqs.(15--16) with a precision of 5 to 10\% for $M_{Z'}=1$ TeV.
This precision deteriorates as the $M_{Z'}$ increases and for 2 TeV,
the error is of the order of 100\%.
The $Z'$ mass and the absolute value of the couplings $g_2,g_{12}$ can be
determined only if the $Z'$ is produced as a resonance, and therefore
calls for higher--energy colliders if the $Z'$ is very heavy. However,
these parameters could be first determined at the LHC as we will discuss
now.

\bigskip

{\it\noindent 2.5 Comparison with the LHC}
\vglue 0.2cm
\baselineskip=14pt

At LHC, new gauge bosons can be searched for by looking at a peak in the
invariant mass spectrum of lepton pairs. The mass reach will depend on
the luminosity of the machine, the specific model from which the $Z'$
originates and since the search relies on the leptonic branching ratio, on
the low energy particle content of the model. At a c.m. energy of 14 TeV
with a luminosity of 100 fb$^{-1}$, requiring 10 $\ee$ and $\mu^+
\mu^-$ events, masses up to 4.5 TeV can be probed if the $Z'$ decays
only into standard particles \cite{R2}. Lowering the luminosity by a
factor of 10 reduces the mass reach by a factor of 3. These limits will
be also lowered if one includes the $Z'$ decay into exotic fermions
[which are the {\bf 27} and {\bf $\overline{27}$} of $E_6$, e.g.] or
supersymmetric particles.

The discovery reach of LHC is shown in Tab.3 for the E$_6$ and LR
models [with $g_L=g_R$] for two c.m. energies and two integrated
luminosities. [Note that similar masses can be reached for charged
vector bosons at LHC, while at $\ee$ colliders the mass reach is rather
limited]. As can be seen, the mass reach of LHC with full energy and a
high luminosity [which will need several years of running] is comparable
with the limits discussed previously at a 500 GeV $\ee$
collider\footnote{Of course, while at the LHC the $Z'$ will be produced
as a real state, one can only indirectly prove its existence at $\ee$
colliders if the energy is not high enough. Nevertheless, discovering
``only" a new neutral current might also be, the least to say,
very interesting [c.f. Gargammelle].} for the same high luminosity of
$\sim 100$ fb$^{-1}$.

\begin{center}
\begin{tabular}{|cc||cccc|} \hline
\ $\sqrt{s}$ [TeV] \ & \ $ \int {\cal L}$ [fb$^{-1}]$ \ &
$ \chi$ & $\psi$ & $\eta$ & LR \\ \hline
10 \ &  40 & 3040 & 2910 & 2980 & 3150 \\
14 \ & \ 100 \ & \ 4308 \ & \ 4190 \ & \ 4290 \ & \ 4530 \ \\ \hline
\end{tabular}
\end{center}
\vspace{0.1cm}
\noindent{\small {\bf Tab.3}: $Z'$ mass reach reach at the LHC with 10
$Z' \ra e^+e^- + \mu^+ \mu^-$ events; from Ref.\cite{R2}.}
\vspace*{.3cm}

Due to the difficult environment at hadron colliders, the identification
of the origin of the $Z'$ is limited to masses below 1 to 2 TeV. The
forward backward asymmetry in the muon channel, the ratio of cross
sections in different rapidity bins and rare processes such as $Z' \ra
WW$ and $pp \ra Z'+W/Z$ are the main probes of the $Z'$ nature
\cite{R2}. For a model independent determination of the $Z'$ parameters,
using the same notation as above, four normalized charges can be probed
at the LHC [in addition to the direct measurement of
the $Z'$ mass; the coupling $g_2$ can be determined from the total $Z'$
width, assuming that only decays into standard particles are present]
\begin{eqnarray}
\gamma_L^e= \frac{(g_{L2}^e)^2}{(g_{L2}^e)^2+(g_{R2}^e)^2} \ , \
\gamma_L^q= \frac{(g_{L2}^q)^2}{(g_{L2}^e)^2+(g_{R2}^e)^2} \ , \
U = \frac{(g_{R2}^u)^2}{(g_{L2}^q)^2} \ , \
D = \frac{(g_{R2}^d)^2}{(g_{L2}^q)^2}
\end{eqnarray}
For $M_{Z'}=1$ TeV, $U/D$ can be determined with a precision of typically
20\% and $\gamma_L^e$ with less than 10\%; the error on $\gamma_L^q$ is
large. Recalling that a 500 GeV LC is sensitive to the normalized charges
(15) and the ratio $g_{Z'}^2/M_{Z'}^2$, it is only the combination of the
information from the LHC and a 500 GeV LC which would allow the model
independent determination of all the parameters of the $Z'$. Therefore,
the LHC and a 500 GeV linear collider are complementary. Increasing
the energy of the LC will increase not only the $Z'$ mass reach but also
the precision with which different models are discriminated and the $Z'$
couplings are measured. The ultimate tests will be of course made
only when a LC with a c.m. energy comparable to the $Z'$ mass will be
available; one would then produce the $Z'$ as a resonance and dissect
the $Z'$ boson to gain information on the physics at the Grand
Unification scale.

\newpage

{\bf\noindent 3. New Matter Particles}
\vglue 0.2cm
\baselineskip=14pt

{\it\noindent 3.1 New Fermions}
\vglue 0.2cm
\baselineskip=14pt

The new fermions predicted by extended gauge models are exotic with
respect to their transformation  under the SM group \cite{F1}. Contrary to
sequential 4th generation heavy fermions [with the neutrino having a
right-handed component for its mass to be generated in a gauge invariant
way] exotic fermions have the usual lepton and baryon quantum numbers
but non--canonical ${\rm SU(2)_L \times U(1)_Y}$  quantum numbers, e.g. the
left--handed components are in weak isosinglets  and/or  the
right--handed components in weak isodoublets.

Except for singlet neutrinos, the new fermions couple to the photon
and/or to the electroweak gauge bosons $W/Z$ [and for heavy quarks, to
gluons] with full strength; these  couplings allow for pair production
with practically unambiguous cross sections. If they have
non--conventional quantum numbers, the new fermions will mix with their
SM  partners. This mixing will give rise to new currents which determine
the decay  properties of the heavy fermions and allow for their single
production. If the  mixing between different generations [which induces
FCNC at tree--level] is  neglected, the mixing pattern simplifies. The
few
remaining angles are restricted by LEP and low energy experiment data to
be smaller than ${\cal O}(0.05-0.1)$ \cite{F1}. Note that LEP1 sets bounds of
order $\sim M_Z/2$ on the masses of  these particles [stronger mass
bounds from Tevatron can be set for  quarks]; masses up to $M_Z$ might
be probed at LEP2.

The heavy fermions decay through mixing into massive gauge bosons plus
their ordinary light partners, $F \ra fZ/ f'W$ \cite{F2,F3,F4}. For masses
larger than
$M_W(M_Z)$ the vector bosons will be on--shell. For small mixing angles,
$\zeta<0.1$, the decay widths are less than 10 MeV (GeV) for $m_F= 0.1
(1)$ TeV. The  charged current decay mode is always dominant and for
$m_F \gg M_Z$, it has a branching fraction of 2/3.

\vspace*{7.5cm}

\noindent {\small {\bf Fig.5}: Pair production of mirror and vector
neutral and charged leptons a 1 TeV $\ee$ collider: a) cross sections
and (b) angular distributions; from \cite{F5}.}

If their masses are smaller than the beam energy, the new leptons can be
pair produced in $\ee$ collisions, $\ee \ra L\bar{L}$, through
$s$--channel gauge boson exchange \cite{F2,F3,F4}. The cross sections are of
the order
of the point--like  QED cross section and therefore,  are rather large.
They  are displayed in Fig.~5a for mirror [they are the same for
sequential] and vector isodoublet leptons at a c.m. of 1 TeV. For a
luminosity of 100 fb$^{-1}$ one expects $10^{3}$--$10^{4}$ events.
Because of their clear signatures [$\ee ZZ$ and $\ee WW$ final states
for charged and neutral leptons respectively], the detection of these
particles is straightforward in the clean environment of $\ee$
colliders, and masses very close to the kinematical limit can be probed
\cite{F5}.

The angular distributions are shown in Fig.5b, and one notes that they
are symmetric for vector leptons leading to $A_{\rm FB}=0$; for mirror
fermions $A_{\rm FB}$ is sizeable and has the opposite sign as for
sequential leptons. [Here, $Z'$ exchange is neglected.]
The total cross sections, angular  distributions and
the polarization of the final particles allow to  discriminate
between
different types of fermions. Charged leptons  can also be
pair--produced at $\gamma\gamma$ colliders, $\gamma \gamma \ra L^+ L^-$,
and for relatively small masses, the  cross sections can be larger than
in the $\ee$ mode.

Note that right--handed neutrinos in LR models can also be produced
in pairs through the exchange of heavy $Z'$ and $W'$ bosons if the
masses are not too large. For $M_{Z'}=M_{W'}=1.5$ TeV, the cross
sections are of the order of a few fb at $\sqrt{s}=500$ GeV
sufficiently below the kinematical threshold \cite{F3}.

For not too small mixing angles, one can also have access to the new
fermions via single production in association with their light partners
\cite{F2,F3,F4}.
The rate for this type of process is more model dependent but can
substantially increase the reach of a given accelerator. In $\ee$
collisions, this proceeds only via $s$--channel $Z$ exchange  [if the
$Z'$ is too heavy] in the
case of  quarks and second/third generation leptons, leading to small
rates. For the first generation leptons, however, one has additional
$t$--channel exchanges [$W$ channel for $N$ and $Z$ channel for $E$]
which increase the cross sections by several orders of magnitude at
high energies.

The cross section for left-- and right--handed neutral and charged
leptons is shown in Fig.~6a  at a c.m. energy of 1 TeV for mixing angles
$\zeta_{L,R}=0.1$. As can be seen, the cross sections are very large
especially for $N_L$ where they can reach the picobarn level. For the charged
leptons they are one order of magnitude smaller, a consequence of the
smaller NC couplings compared to the CC couplings. For smaller mixing
angles the rates have to be scaled down correspondingly; however, even
for $E$ and $N_R$ requiring 10 events with a luminosity $\int {\cal
L}=100$ fb$^{-1}$, one can probe $\zeta$ values one order of magnitude
smaller for $m_L=800$ GeV \cite{F5}.

The angular distributions are shown in Fig.6b: it is clear that one can
easily distinguish between neutrinos with left- and right-handed
couplings and of Dirac and Majorana nature. A further distinction
[especially for $E_{L,R}$ which have the same cross section and
distribution, a consequence of the fact that the vector coupling of the
electron to the $Z$ boson is small] can be made with the longitudinal
polarization of the initial beams, and also with the polarization of the
final leptons.

\newpage

\vspace*{8.5cm}

\noindent {\small {\bf Fig.6}: Single production of left-- and
right--handed leptons a 1 TeV $\ee$ collider: a) cross section as a
function of $m_L$ and (b) angular distribution for $m_L=750$ GeV; the
mixing angles are taken to be $\zeta_{L,R}=0.1$; from \cite{F5}.}
\vspace*{0.4cm}

To fully reconstruct the heavy lepton masses, the best signals consist
of an $\ee$ pair and two jets for the charged lepton and an $e^\pm$, a
pair of jets and missing momentum for the neutral lepton; the branching
ratios are $23\%$ and 43\% respectively. In the case of $E$, the main
backgrounds are: $ e^+ e^-\to e^+e^-Z \to \ee jj$, $ e^+ e^- \to Z Z$, $
e^+ e^- \to t \bar{t} \to W^+W^- jj$ and $ \gamma \gamma \to  e^+ e^-q
\bar{q}$. In the case of $N$, the backgrounds are: $ \ee \to e \nu W$, $
\ee \to WW \to e^\pm \nu_ jj$ and $ \gamma \gamma \to e (e)  q  \bar{q}
$. These backgrounds can be eliminated or reduced by applying the cuts
shown in Table 4. After these cuts, no events from heavy flavor
production or from the $\gamma \gamma$ backgrounds would survive; the
backgrounds from vector boson production can be suppressed to a very low
level, while those from single $W/Z$ production can be a bit higher.

A full simulation \cite{F4} of the signal and backgrounds has been
performed using PYTHIA, for a model detector [an upgraded LEP detector]
to quantify the discovery limits that can be obtained. This simulation
was done assuming a c.m. energy of 500 GeV and an integrated luminosity
of 50 fb$^{-1}$. The signal and background cross sections after applying
cuts are shown in Tab.4 [note that at $\sqrt{s}=500$ GeV, the cross
sections are practically the same at 1 TeV, because the dominant
contribution comes from the $t$-channel exchange] for heavy leptons with
masses of 350 GeV and with $\zeta=0.05$ for $E$ and $\zeta_L=0.025$ for
$N$. For these $\zeta$ and mass values, the signal peaks stand out
clearly from the background events. For $m_L=350$ GeV and requiring that
the signal over the square-root of the background is larger than unity,
one can probe mixing angles down to $\zeta \sim 0.005$ for neutral
leptons and $\zeta \sim 0.03$ for charged leptons. For $m_E \sim 450$
GeV, only slightly smaller $\zeta$ values can be probed. The situation
is much more favorable for $N_L$, the cross section being one order of
magnitude larger. At $\sqrt{s}=1$ TeV, these numbers for $m_L$ and
$\zeta^2$ can be improved by a factor of two.

\begin{small}
\begin{table}[htbp]
\centering
\begin{tabular}{|c||c|c|c|} \hline
Process & $E^\pm e^\mp$ &  $ e^+ e^- Z $ &   $ Z  Z $  \\ \hline
$ \sigma$ [fb] &    9.5 & 4960 &  615 \\ \hline
$ \times $ B.R. & 2.19 & 3470 & 28.8 \\ \hline
one $\ee$ pair & 1.74 & 93.0 & 23.0 \\
$ 330 < M_E < 370$ & 1.56 & 11.7 & 5.30 \\
$  85 < M_Z < 105$ & 1.41 & 5.84 & 2.87 \\
$ |M_{ll}-M_Z|>12$ & 1.39 & 5.18 & 1.02 \\
$\cos\theta_{ll}<0.5$ & 1.33 & 4.32 & 0.56 \\
$ f( M_E, \cos \theta_{Z})$ & 1.30 & 1.90 &  0.43 \\
 kinem. cuts & 1.30 & 1.55 & 0.39 \\ \hline
\end{tabular}
\hspace*{1mm}
\begin{tabular}{|c||c|c|c|} \hline
Process  &$ \bar{N} \nu $ & $ e \nu W $& $  W  W $  \\ \hline
 $ \sigma$ [fb]                             &    490                     &
8610             &   2600       \\ \hline
 $ \times $ B.R.
                                       &    13.7                    &
5823             &   1140       \\ \hline
  one $e$                              &    13.2                    &
198              &    883       \\
$ 330 < M_N < 370$ &    12.5                    &
11.9             &    100       \\
$  70 < M_W <  90$ &    12.3                    &
10.3             &     70.3     \\
$  M_{l\nu} > 120 $&    11.8                    &
10.0             &     7.93     \\
$ \cos \theta_{l\nu} < 0.5 $              &    11.7                    &
10.0             &     7.80     \\
$ f( M_N, \cos \theta_{Z})$ &    11.7                    &
10.0             &     7.80     \\
 kinem. cuts                           &    11.7                    &
10.0             &     4.13     \\ \hline
 \end{tabular}
\end{table}
\end{small}

\nn {\small {\bf Tab.4}: Cross sections for heavy lepton single production and
for the main backgrounds at $\sqrt{s}=0.5$ TeV after successive applications
of cuts; $m_L=350$ GeV and $\zeta= 0.025(0.05)$ are chosen for $N(E)$ and the
masses are in GeV.}
\vspace*{0.3cm}

Note that heavy fermions cannot be produced singly at $\gamma
\gamma$ colliders [at least in a $2\ra 2$ process]; heavy neutral and
charged leptons can be produced in $e\gamma$ collisions in association with
massive gauge bosons \cite{F6}, however only smaller masses can be probed
and the rates are not much larger than in $\ee$ collisions.

Heavy leptons can also be searched for indirectly. A prominent example
is the search for Majorana neutrinos in the reaction $e^-e^- \ra W^-W^-$
which is similar in nature to neutrinoless $\beta \beta$ decay. This
process has been discussed in \cite{F7,F8,F9}. In the breaking of $E_6$
down to the SM group through the SO(10) chain, one could assume that
the right--handed $W$ bosons [as well as the doubly charged Higgs
bosons] are very heavy but the two additional isosinglet neutrinos of
$E_6$ have masses in the ${\cal O}$ (TeV) range \cite{F8}.
It has been shown that this scenario is still open
to experimental detection through the process $e^- e^- \ra W^- W^-$
where at least the lightest of the two Majorana neutrinos is exchanged.
For energies well above the mass of the $W$ but below that of the
Majorana neutrino [where no doubly charged Higgs boson exchange is
needed for unitary reasons], the cross section is proportional to $s^2$
and therefore can be observed even for very small values of the
suppressing mixing angles \cite{F8}. The use of polarization enhances
the rates and the spectacular back--to--back $W$ pair allows effective
background suppression, and even a moderate signal may lead to
convincing discovery. It is not yet clear whether constraints from
neutrinoless $\beta \beta$ decay are not in conflict \cite{F9} with
this scenario, though. A more detailed analysis is required.

Finally, heavy quarks can also be pair produced in $\ee$ and $\gamma
\gamma$ collisions for masses up to the beam energy. However, they can
be best searched for at hadron colliders where the production processes,
$gg/ q \bar{q} \ra Q \bar{Q}$, give very large cross sections: at LHC
with $ \sqrt{s}=14$ TeV and a luminosity of 10 fb$^{-1}$ quark masses up
to 1 TeV can be reached \cite{F5}. However, and $\ee$ linear collider would be
needed to study their properties, a situation similar to the one of the
top quark.

\newpage

{\it\noindent 3.2 A Test of Superstrings inspired $E_6$ models}
\vglue 0.2cm
\baselineskip=14pt

In superstrings inspired $E_6$ models, the superpartners of the third
generation doublet $L_3, \bar{L}_3$ and singlet $n_3$ fields provide the
Higgs sector of the theory; while the doublets [with vacuum expectation
values $<\tilde{L}_3>=v$ and $<\tilde{\bar{L}}_3>=\bar{v}$] give masses
to the $W/Z$ bosons, the singlet [with a vev $<\tilde{n}_3>=x$] gives
mass to the $Z'$ [if the vev $x$ is large, the $Z'$ will be very
heavy and the model in principle reduces to the minimal supersymmetric
model at low energies]. The fermionic third generation will then mix
with the gauginos to form the neutralino fields. The first two
generations of leptons $L_{1,2}, \bar{L}_{1,2}$ and $n_{1,2}$ obtain
their masses [the SO(10) neutrino acquire mass through a different
mechanism] from the superpotential $W_{\rm lep}= \sum \lambda_{ijk} L_i
\bar{L}_j n_k$ with $i,j,k=1,2,3$. Assuming that $\lambda_{33i}=
\lambda_{3i3}= \lambda_{i33} =0$ these leptons will not mix with the
gaugino--higgsino sector. The previous superpotential leads to a mass
matrix, which under the assumption that no singlet scalar gets a vev
much larger than $10^{9}$ GeV, gives a very strong constraint on the
mass of two neutral leptons. This will provide a decisive test of these
models as has been discussed by Drees \cite{F10} and is summarized
below.

While some entries of the mass matrix are proportional to the masses of
the charged leptons which can be made very large since they are $\propto
x$, other entries must be of ${\cal O}(100)$ GeV or less if the
$\lambda$ couplings have no Landau pole at low energies. Since
the determinant of the matrix will be proportional to the square of
the product of the three vev's, four eigenvalues are large [$\propto
m_{L^\pm_i}$ with $i=1,2$] and two eigenvalues must be small
\begin{eqnarray}
{\rm det} {\cal M} \sim ( x v \bar{v} )^2 \ \ \ \rightarrow \ \ \
m_{ L_{1,2}^0 } \sim \lambda^2 v \bar{v} / m_{L^\pm_i}
\end{eqnarray}
The upper bound on the light neutral lepton mass will therefore decrease
as the charged lepton masses increase. Therefore, the negative search of
charged heavy leptons or associated $L_1^0 L_3^0$ production in $\ee$
collisions will place an upper bound on the mass of the light leptons.
For instance at a 1.5 TeV collider, the mass of the lightest lepton
should be smaller than 30 GeV for $\lambda_{\rm max}=0.85$ [so that the
model remains weakly interacting up to high scales] and $\bar{v} /v=1$
[stronger bounds are obtained for higher values] in the case of a
negative search. This means that, since the exotic leptons have R--odd
parity, $L_1^0$ will be the lightest supersymmetric particle. If
R--parity is conserved, $L_1^0$ must be therefore absolutely stable and
will contribute to the density of Big Bang relics and might overclose
the universe. Requiring that the annihilation cross section for
$L_{1,2}^0$ pairs is large enough not to lead to cosmological problems,
the couplings of $L_{1,2}^0$ to the light Higgs boson should be large,
and since $m_h >2m_{L}$ the decay $h \ra L_1^0 L_1^0$ is kinematically
possible, the branching branching ratio of invisible Higgs decays will be
larger than 50\%.

Thus, either one discovers a light Higgs [in SUSY models the lightest
boson has a mass smaller than $\sim$ 150 GeV] with a large invisible
branching at a 300 GeV $\ee$ collider, or find a heavy lepton in the
channel $\ee \ra L_1^0L_3^0$ at a 1.5 TeV $\ee$ collider. Otherwise,
the model would be ruled out. Therefore, $\ee$ colliders could provide a
decisive test of superstring inspired models.

{\it\noindent 3.3 Difermions}
\vglue 0.2cm
\baselineskip=14pt

In addition to the usual couplings to gauge bosons, difermions
\cite{S12} have
couplings to fermion pairs which determine their decays [here also one
can neglect the couplings between different generations to prevent FCNC
at tree-level].  These couplings are a priori unknown.  In the case of
leptoquarks (LQ) for example, a systematic description of their quantum
numbers and interactions can be made by starting from an effective
lagrangian with general SU(3)$\times$SU(2)$\times$U(1) invariant
couplings and conserved  B and L numbers. This leads to the existence of
5 scalar and 5 vector LQ's  with distinct SM transformation properties.
In general, present data constrain  difermions to have masses larger
than 50--150 GeV.

Leptoquarks can be produced in pairs at $\ee$ colliders through gauge
boson exchange; significant $t$-channel quark exchange can be present in
some channels if the quark-lepton-LQ couplings are not too small.
Depending on  the charge, the spin and isospin of the LQ, the cross
sections can vary  widely [at $\sqrt{s}=500$ GeV \cite{F11} between $\sim 10$
fb
and $\sim 3.5$ pb]. Through the signatures of 2 leptons plus 2 jets, these
states are accessible for masses smaller than the beam energy.  The
study of the various final states and the angular  distributions would
allow the determination of the quantum number of the LQ's as in the case
of exotic fermions \cite{F5}. LQ's can also be  pair produced in $\gamma
\gamma$
collisions; depending on the LQ charge, the cross sections can be much
larger or much smaller than for charged leptons.

Single production of scalar and vector leptoquarks can also take place
in the $\ee$, $e^- e^-$, $e\gamma$ and $\gamma \gamma$ modes of the collider
\cite{S13}. The kinematical reach is thus extended to $\sqrt{s}$ but
the production rates are suppressed  by the unknown LQ coupling to
quark--lepton pairs. At a 1 TeV $\ee$ collider with $\int{\cal L}=60$
fb$^{-1}$, one reaches masses close to $\sqrt{s}$ [i.e. 1 TeV for $\ee$
and $e^- e^-$, $\simeq 0.9$ TeV  for $e \gamma$ and $\simeq 0.8$ TeV for
$\gamma \gamma$] \cite{F5}.

Since they are strongly interacting particles, LQ's can be produced at
hadron colliders with very large rates. At LHC with 100 fb$^{-1}$ the search
reach for scalar/vector LQ's is 1.4/2.2 TeV if one
assumes a branching fraction of unity for the $eejj$ final state
\cite{F5}. Therefore,
LQ's can be best searched for at the LHC; however, $\ee$ colliders could
provide very important informations on their properties.

Dileptons can be pair produced in $\ee/\gamma  \gamma \ra X^{++}X^{--}$
and masses up to $\sqrt{s}/2$ can be probed; the rates [especially in
$\gamma \gamma$ collisions because of the charge]  are very large and
the signatures [four leptons] are  spectacular. Dileptons can also be
singly produced in the four modes of the collider. In particular, in
the $e^- e^-$ mode di--electrons can be produced as  $s$--channel
resonances. At a 1 TeV collider, scalar and vector dileptons  can
be observed up to masses of $\sim 0.9$ TeV in the $e\gamma$ mode even
for couplings to lepton pairs as small as $10^{-3}$ the electromagnetic
coupling; in the $\ee$ and  $\gamma \gamma$ modes, dileptons can be
observed for couplings an order of  magnitude larger.

Finally, diquarks can be pair produced in $\ee$ and $\gamma \gamma$
collisions for masses smaller than $\sqrt{s}/2$ with appreciable rates,
with a signal consisting of an excess of 4 jets events \cite{F5}. They can be
also
pair produced at hadron colliders, either in pairs or singly [for the
first generation]  if the couplings to quark pairs is not too small.
However, since the signals consist only in jets, the large QCD
backgrounds might be a problem.

\newpage

{\bf\noindent 4. Summary}
\vglue 0.2cm
\baselineskip=14pt

We have summarized the discovery potential of high--energy $\ee$ linear
colliders with respect to the new matter particles and the new gauge
bosons predicted by gauge extensions of the Standard Model.

If the energy can be raised high enough, the $\ee$ collider will operate
as a $Z$' factory; the event rates will be very high and the properties
of the $Z'$ can be studied in great details. A heavy $Z'$ boson, even if
its mass is substantially larger than the available center of mass
energy, will manifest itself through its propagator effects in the
process $\ee \ra$ fermions, producing potentially sizeable effects on
the observables $\sigma^{\rm lept} , R= \sigma^{\rm had}/\sigma^{\rm
lept} , A_{\rm FB}^{ \rm lept}$ and if longitudinal polarization is
available $A_{\rm LR}^{\rm lept}$ and $A_{\rm LR}^{\rm had}$. Masses up
to 6 times the c.m. energy of the collider can be probed for the
expected luminosities. If a $Z'$ with mass below 3 TeV is discovered at
LHC, even a 500 GeV $\ee$ collider would give valuable contributions to its
detailed investigation by allowing the distinction between different
classes of models and the determination of the model parameters. The two
types of colliders would then provide complementary information.

$\ee$ colliders are well suited machines for the search of new leptons.
These particles can be produced with large rates if their masses are
smaller than the beam energy. They can also be singly produced in
association with their standard light partners if the mixing angles are
not prohibitively small; one can then reach masses close to the total
energy of the collider. The signatures have clear characteristics so that
the detection of these particles should not be difficult in the clean
environment of $\ee$ colliders. Since they are strongly interacting
particles, quarks, leptoquarks and diquarks will be produced at the LHC
with very large rates. However, because of the difficult hadronic jet
background, the signals would be hard to analyze in detail. $\ee$
colliders would provide the ideal framework for highly precise analyses
of the properties of these new exotic particles if they are found at
the hadron colliders.

\bigskip

{\bf\noindent Acknowledgements:}
\vglue 0.2cm
\baselineskip=14pt

I thank the organizers of the Workshop for their invitation, their
support and the very nice and stimulating atmosphere of the meeting.
Discussions with M. Drees, C. Heusch, A. Leike, P. Minkowski and C.
Verzegnassi are gratefully acknowledged.

\bigskip

{\bf\noindent References}
\vglue 0.2cm

\end{document}